# The Decay of Two Adjacent Sunspots Associated with Moving Magnetic Features

Yang Peng[1,2], Zhike Xue[1,3], Zhongquan Qu[1,2], Jincheng Wang[1,3], Zhe Xu[1,3], Liheng Yang[1,3], and Yian Zhou[1]
[1] Yunnan Observatories, Chinese Academy of Sciences, Kunming 650216, People's Republic of China; zkxue@ynao.ac.cn
[2] University of Chinese Academy of Sciences, Beijing 100049, People's Republic of China
[3] Yunnan Key Laboratory of Solar Physics and Space Science, Kunming 650216, People's Republic of China

## Abstract

The relationship between the decay of sunspots and moving magnetic features (MMFs) plays an important role in understanding the evolution of active regions. We present observations of two adjacent sunspots, the gap between them, and a lot of MMFs propagating from the gap and the sunspots' outer edges in NOAA Active Region 13023. The MMFs are divided into two types based on their magnetic field inclination angle: vertical ($0° < \gamma < 45°$) and horizontal ($45° \leqslant \gamma < 90°$) MMFs (V-MMFs and H-MMFs, respectively). The main results are as follows: (1) the mean magnetic flux decay rates of the two sunspots are $-1.7 \times 10^{20}$ and $-1.4 \times 10^{20}$ Mx day$^{-1}$; (2) the magnetic flux generation rate of all MMFs is calculated to be $-1.9 \times 10^{21}$ Mx day$^{-1}$, which is on average 5.6 times higher than the total magnetic flux loss rate of the sunspots; (3) the magnetic flux of V-MMFs (including a pore separated from the sunspots) is 1.4 times larger than the total lost magnetic flux of the two sunspots, and in a later stage when the pore has passed through the reference ellipse, the magnetic flux generation rate of the V-MMFs is almost the same as the magnetic flux loss rate of the sunspots; and (4) within the gap, the magnetic flux of V-MMFs is one-third of the total magnetic flux. Few V-MMFs stream out from the sunspots at the nongap region. All observations suggest that MMFs with vertical magnetic fields are closely related to the disintegration of the sunspot, and most of the MMFs from the gap may originate directly from the sunspot umbra.

*Unified Astronomy Thesaurus concepts:* Sunspots (1653); Solar magnetic fields (1503); Solar photosphere (1518)

*Supporting material:* animation

## 1. Introduction

Sunspot decay makes a significant contribution to the evolution of the solar large-scale magnetic field (Wallenhorst & Howard 1982; Sheeley et al. 2017). Photometric and magnetic decay are two different patterns of the decay (Martínez Pillet 2002), and their onsets are not synchronous (Li et al. 2021). Two decay models are proposed to explain sunspot decay: turbulent diffusion and turbulent erosion models. The turbulent diffusion model implies that sunspot decay is independent of the size and perimeter of the sunspot, and magnetic flux loss occurs everywhere in the sunspot (Meyer et al. 1974; Martínez Pillet 2002). By contrast, the turbulent erosion model indicates that sunspot decay is correlated with the perimeter and erodes from the boundary (Simon & Leighton 1964; Petrovay & Moreno-Insertis 1997; Litvinenko & Wheatland 2015).

The area decay of a sunspot, as an important magnetic activity indicator, has been extensively studied. Bumba (1963) obtained a linear area decay rate of −4.2 millionths of a solar hemisphere (MSH, 1 MSH = 3.32 Mm$^2$) day$^{-1}$ for recurrent sunspots. Subsequently, Martinez Pillet et al. (1993) determined a mean decay rate of −12.1 MSH day$^{-1}$ for sunspot groups. Comparing with their results, Li et al. (2021) found that the decay rates of $\alpha$-configuration sunspots range from −7.75 to −23.81 MSH day$^{-1}$. Some research has suggested that the decay rates depend on the area and polarity of the sunspot (Gómez et al. 2014; Norton et al. 2017; Muraközy 2020), and that they satisfy the lognormal distribution (Martinez Pillet et al. 1993). Linear (Robinson & Boice 1982; Solanki 2003) and quadratic (Moreno-Insertis & Vazquez 1988; Martinez Pillet et al. 1993) decays are two main patterns of sunspot decay, described by the turbulent diffusion and turbulent erosion models, respectively.

The magnetic decay of sunspots plays a prominent role in the flux transportation of the solar surface. Numerous studies have suggested that the magnetic flux loss rates of sunspots are around $10^{20}$ Mx day$^{-1}$ (Verma et al. 2012; Sheeley et al. 2017; Li et al. 2021; Peng et al. 2023). The fragmentation and disintegration of the sunspot magnetic flux are associated with light bridges (Vazquez 1973; Louis et al. 2012; Murabito et al. 2021) and moving magnetic features (MMFs) (Kubo et al. 2008a). Li et al. (2023) observed that the formation of a light bridge may accelerate the decay of the penumbra within a decaying sunspot. A number of researches have indicated that the magnetic field of the penumbra becomes more vertical during sunspot decay (Li et al. 2021; Peng et al. 2023). Wang et al. (2004) observed a rapid penumbral decay after three solar flares. They suggested that a part of the penumbral magnetic field converts into an umbral field, and the penumbral field becomes more vertical. Watanabe et al. (2014) investigated the formation and decay of a penumbra around a pore and found that the penumbral magnetic field also becomes more vertical during the decay of the penumbra. They proposed that the recovery of a dark umbral area may be responsible for the change of the penumbral magnetic field inclination angle.

MMFs are small-scale magnetic structures streaming out radially from sunspots. MMFs exhibit an average size of less than 2″ and their horizontal velocity ranges from 0.1 to 2 km s$^{-1}$ (Hagenaar & Shine 2005; Kaithakkal et al. 2017; Li et al. 2019). There are three types of MMFs (Shine & Title 2000). Type I MMFs are bipolar, and unipolar MMFs with the same or opposite polarity to their parent sunspots are described as Type II or III MMFs, respectively. The relationship between the magnetic flux loss of a decaying sunspot and its surrounding MMFs is still mysterious. Criscuoli et al. (2012) tracked six Type II or III MMFs and found their magnetic field strength ranges from 500 G to 1700 G, and the inclinations are all vertical. A mean magnetic field strength of 1500 G for unipolar MMFs was obtained by Li et al. (2019), and the magnetic fields of a part of these MMFs are highly inclined. The connection between the origin of MMFs and penumbral filaments has been studied (Zhang et al. 2003; Sainz Dalda & Martínez Pillet 2005; Cabrera Solana et al. 2006). Thomas et al. (2002) and Weiss et al. (2004) suggested that Type II MMFs correspond to magnetic flux tubes detached from the vertical component of the penumbral uncombed structure and being transported outward by convection. They are responsible for the disintegration of sunspots. Unlike Type II MMFs, Type I and III MMFs are formed due to the intersection of the horizontal field extended from the penumbra, and do not contribute to the magnetic flux loss of sunspots. Martínez Pillet (2002) found that the magnetic flux generation rates of MMFs are 3–8 times higher than the flux-loss rates of the sunspots, implying that most MMFs cannot be related to the decay of sunspots. Kubo et al. (2007) found the magnetic fields of MMFs that are situated in extrapolated lines of the penumbral spine are more vertical to the solar surface, and only isolated vertical MMFs are devoted to the disappearance of sunspots. Li et al. (2019) observed 268 unipolar MMFs and suggested that both Type II and Type III MMFs contribute to the evolution of the sunspot. Many MMFs also exist near pores, and thus the penumbra may not be a requisite for the generation of MMFs (Deng et al. 2007; Criscuoli et al. 2012; Verma et al. 2012).

In this paper, we present the decay of two adjacent sunspots associated with MMFs in NOAA Active Region (AR) 13023 from 2022 May 27 to 2022 June 2. We concentrate on the relationship between the magnetic flux decay of the two sunspots and the transportation of the magnetic flux by MMFs. Observations and data reductions are described in Section 2. Section 3 analyzes the results of data reduction. Section 4 summarizes conclusions and discusses these conclusions.

## 2. Observations and Data Reductions

Table 1 shows the information of NOAA AR 13023's evolution process. To avoid limb artifacts, the location of the AR is limited to $\pm 40°$ in central meridian distance in our study. Therefore, we define the start and end times as the moments when the AR is first inside of $-40°$ and first outside of $40°$ from the central meridian. The AR number, date, and location interval of the evolution process are listed in Table 1. The continuum images and vector magnetic field data are selected from the series hmi.sharp_cea_720s. This series is Space-weather HMI Active Region Patches data taken by the Helioseismic and Magnetic Imager (HMI; Schou et al. 2012) on board the Solar Dynamics Observatory (SDO; Scherrer et al. 2012). Its temporal cadence and spatial resolution are 720 s and 1″, respectively. The vector magnetic field data consist of AR indices of cylindrical equal-area (CEA) projections of $B_\phi$, $B_\theta$, and $B_r$. $B_\phi$ and $B_\theta$ represent the $\phi$ (westward) and $\theta$ (southward) components of the CEA projection vector magnetic field, and the radial (out-of-the-photosphere) component of the CEA vector magnetic field is $B_r$ (Hoeksema et al. 2014). The horizontal magnetic field ($B_t$) and the inclination angle ($\gamma$) of the magnetic field are calculated by Equations (1) and (2),

**Table 1**
Sunspot Information

| AR Number | Start Date | Start Position | End Date | End Position |
|---|---|---|---|---|
| NOAA 13023 | 20220527 16:00 UT | S14°E40° | 20220602 21:00 UT | S14°W40° |

respectively. The vertical magnetic field $B_z$ here is expressed by $B_r$.

$$B_t = \sqrt{B_\phi^2 + B_\theta^2}, \quad (1)$$

$$\gamma = \arctan\left(\frac{B_t}{|B_r|}\right)\frac{180°}{\pi}. \quad (2)$$

Equation (3) gives the unsigned magnetic flux ($\Phi$), where $dA$ is the differential area:

$$\Phi = \int |B_r|\, dA. \quad (3)$$

The umbra and penumbra are identified by the following method. First, the mean continuum intensity ($I_0$) of the solar quiet region is calculated. Second, the umbral and penumbral regions satisfy the following criteria, $I_{\mathrm{umbra}} \leqslant 0.5I_0$ and $0.5I_0 < I_{\mathrm{penumbra}} \leqslant 0.9I_0$, where $I_{\mathrm{umbra}}$ and $I_{\mathrm{penumbra}}$ are the continuum intensities of the umbra and penumbra, respectively. Many studies show that the inner boundary threshold of the penumbra lies in $0.5I_0$–$0.6I_0$ (Benko et al. 2018; Jurčák et al. 2018; Schmassmann et al. 2018; Löptien et al. 2020; García-Rivas et al. 2021; Li et al. 2023), and the outer boundary threshold of the penumbra is around $0.9I_0$ (Li et al. 2021, 2022, 2023; Murabito et al. 2021). We determine the umbral boundaries for each image using the fixed threshold of $0.5I_0$, and the determined umbral boundaries are well adapted to the entire sunspot evolution during the observation. Moreover, the hole-filling and region-growing methods rule out wrongly identified regions. The region-growing method groups neighboring pixels of the seed point according to the specified threshold (Gonzalez & Woods 2002). It efficiently identifies the different components of sunspots and excludes interference of surrounding pores. The hole-filling technique is an image-processing method that fills in holes in the connected domain (Gonzalez & Woods 2002); it fills in misrecognized regions in the penumbra. Finally, the boundaries of the umbra and penumbra are superposed on the continuum and vector magnetic field maps, respectively. The left panel of Figure 1 displays the penumbral inner and outer boundaries of the two sunspots obtained based on the above method.

The Differential Affine Velocity Estimator for Vector Magnetograms (DAVE4VM) method (Welsch et al. 2007; Schuck 2008) calculates the horizontal velocity field, which is used to estimate the magnetic flux of the MMFs ($\Phi_{\mathrm{MMFs}}$) through curve $L$ (an ellipse or arc) in $\triangle t$ by Equation (4). $dL$ is the differential length of the curve $L$, and $\triangle t$ represents the temporal cadence of 720 s; $v_\perp$ is the component of the horizontal velocity in the normal direction of $dL$, and $\langle B_z \rangle$ is the

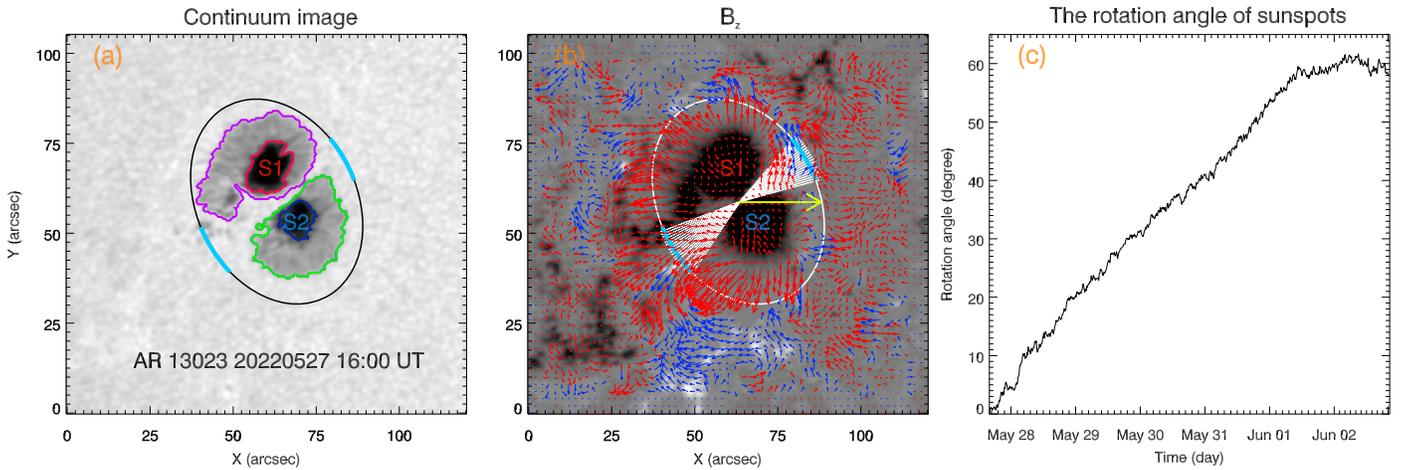

**Figure 1.** Penumbral inner and outer boundaries, and the magnetic flux calculation of the MMFs from the sunspots and gap. Panels (a) and (b) are the continuum image and longitudinal magnetic field ($B_z$) map taken by SDO/HMI on 2022 May 27 at 16:00 UT. The left and right sunspots are marked S1 and S2 in panel (a). The red, blue, magenta, and green contours are the penumbral inner boundaries of S1 and S2 and their penumbral outer boundaries, respectively. The ellipse measures the magnetic flux generated by MMFs from the sunspots. The blue arcs measure the magnetic flux flowing out of the gap. Panel (b) displays the distribution of plasma velocities based on the method DAVE4VM. The blue and red arrows represent the motion directions for positive and negative plasma. Their length represents the velocity magnitude. The fan-shaped region measures the velocity and direction of the MMFs from the gap. The rotation angle evolution of the sunspots is shown in panel (c). An animation of the SDO/HMI observations of NOAA AR 13023 is available online. The animation runs from 2022 May 27 at 16:00 UT to 2022 June 2 at 20:48 UT.

(An animation of this figure is available.)

mean vertical magnetic field strength on $dL$.

$$\Phi_{\mathrm{MMFs}} = \oint_L v_\perp \cdot \triangle t \cdot dL \cdot \langle B_z \rangle. \qquad (4)$$

To calculate the flux of MMFs, an ellipse curve is determined by the following method. The centroid of the two sunspots is the ellipse center, and the major axis of the ellipse passes through the centroid of each sunspot and shrinks steadily with the decay of the sunspots so that the ellipse is close to the penumbral outer boundary (see the ellipse in Figures 1(a) and (b)). Two fan-shaped regions (the dotted lines in Figure 1(b)) are determined to track the MMFs from the gap. Their centroids are the same as that of the ellipse and their radii are 20″. It is should be noted that the sunspots rotate counterclockwise around each other, and their rotation angle is displayed in Figure 1(c). The sunspots rotate at a speed of 12°.6 day$^{-1}$ before 11:00 on June 1, and the total rotation angle is 60°. After that, the two sunspots stop rotating. Thus, the ellipse and the fan-shaped regions also rotate. Furthermore, the MMFs (consisting of isolated and nonisolated MMFs) are divided into horizontal ($45° \leqslant \gamma < 90°$) and vertical ($0° < \gamma < 45°$) MMFs based on Kubo et al. (2007).

## 3. Results

Figure 2 displays the evolution of NOAA AR 13023. At the beginning of the observation, this AR contains two sunspots that are close together, and they are both of negative polarity. The Hale class of NOAA AR 13023 is α-type. During the period from May 27 to June 2, the umbra and penumbra of the sunspots shrink (see Figures 2(a1)–(a4)). Furthermore, the areas and magnetic fluxes of sunspots S1 and S2 all decrease with time (see Figure 3), implying that the two sunspots are decaying, and they are separated by a gap (see the green arrows in Figure 2(a1)). Below sunspot S1, in the penumbra, there is a patch whose intensity is the same as that of the umbral boundary. Compared with other regions in the penumbra of S1, this patch has a stronger magnetic field and the inclination of this magnetic field is more vertical (see Figures 2(a1), (b1), and (e1)). In terms of the previous evolution of the sunspots, the patch is split from the umbra of S1; however its continuum intensity is in the penumbra criterion and it is recognized as a part of S1's penumbra. And below sunspot S1, outside the penumbra, a large number of magnetic patches flow out of the sunspots (see the red arrow in Figure 2(d1)). Traces of these magnetic patches can also be clearly seen in the continuum map (see Figure 2(a1)). At 10:48 UT on May 28, a large penumbra in the southeast detaches from sunspot S1 and gradually evolves into a pore (see Figure 2(a2)). Subsequently, the pore gradually moves away from sunspot S1 and decays, eventually disappearing in the continuum and $B_z$ maps. At 4:48 UT on May 30, the region where the penumbra breaks away (marked by the orange arrow in Figure 2(a3)) reforms the penumbra. Meanwhile, we also clearly observe that the two sunspots rotate around the centroid. Before 11:00 on June 1, the two sunspots rotate rapidly, but later, they almost stop rotating. A number of MMFs propagate from the sunspots' outer edge, and the MMFs from the gap are more obvious in the $B_t$ and $B_z$ maps. Most of the MMFs have the same polarity as the sunspots, and a few have positive polarity. At the end of the observation, the two sunspots do not completely decay, and maintain their complete structure, including the umbra and penumbra, but the sunspots are much smaller than they were in the beginning. In the course of the observation, no obvious solar eruption activity occurs in this AR.

### 3.1. The Decay of the Sunspots

Figure 3 displays the area and magnetic flux decay of the two sunspots with time. The area of each sunspot is fitted by a linear formula: $A(t) = A_0 + DA \cdot (t − t_0)$, where $A(t)$ represents the area (unit: MSH) at time $t$ (unit: days), $t_0$ is the start time of the observation, $A_0$ is the value of each fitting contour at time $t_0$, and $DA$ is the decay rate of the area. Figure 3(a1) shows that the umbral area of sunspot S1 first decreases linearly from 20.5 MSH to 14.5 MSH at a decay rate of −3.7 MSH day$^{-1}$,

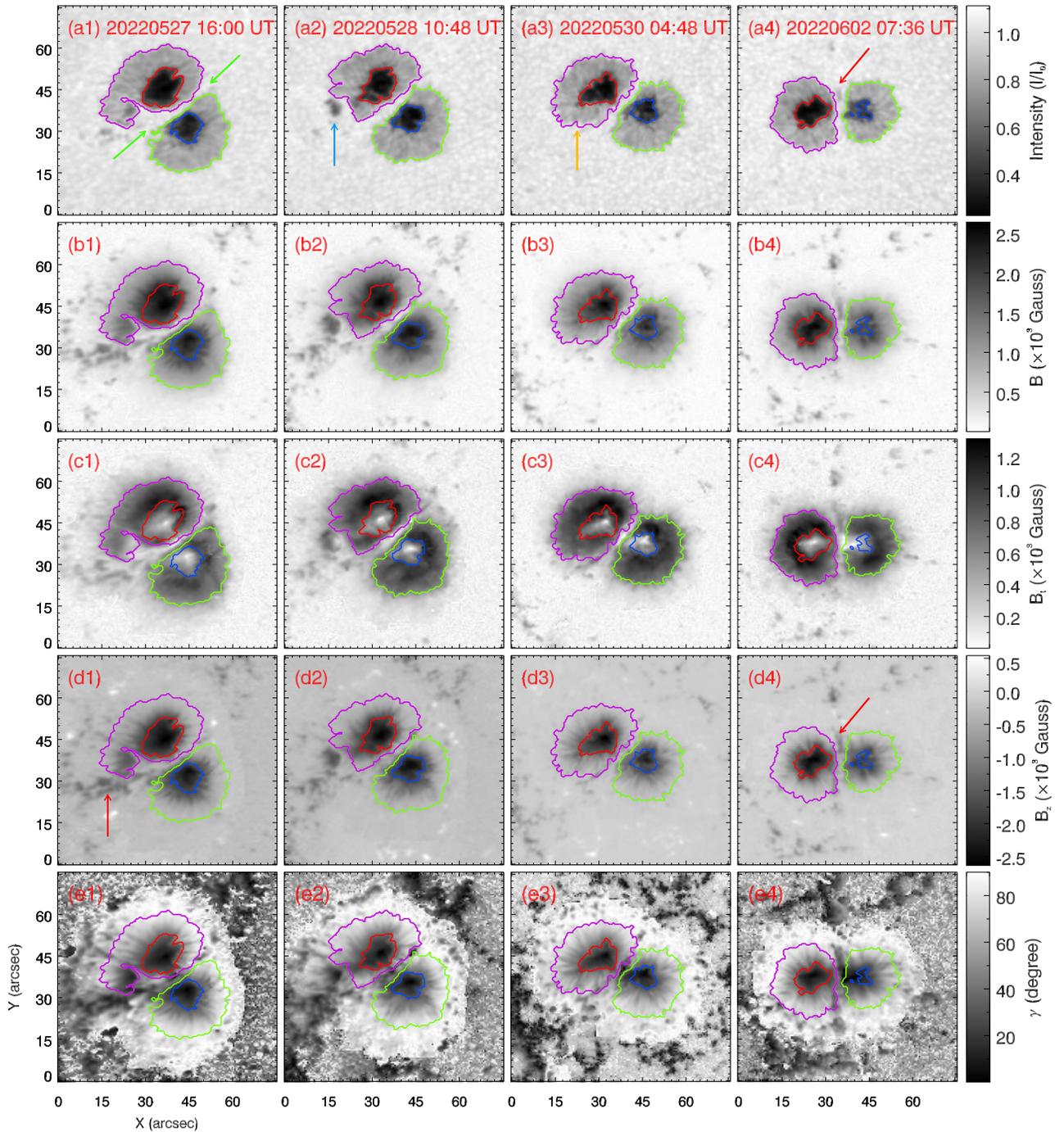

**Figure 2.** Evolution of NOAA AR 13023. The contour colors are the same as those of Figure 1. The uppermost row shows the continuum maps at four moments of the sunspot decay. The second, third, fourth, and fifth rows are the $B$, $B_t$, $B_z$, and $\gamma$ maps at the corresponding moments of the continuum maps in the first row. $B$, $B_t$, $B_z$, and $\gamma$ are the magnetic field strength, transverse magnetic field strength, longitudinal magnetic field strength, and magnetic field inclination angle, respectively. The green arrows mark the location of the gap at the initial time of the observation. The red arrows mark the MMFs from the gap at the initial and ending time of the observation. The blue arrow marks the penumbra detached from sunspot S1. The orange arrow indicates that the region where the penumbra breaks away reforms its penumbra.

and has a slow linear reduction to 12.4 MSH at $-0.5$ MSH day$^{-1}$. In contrast, the umbral area of S2 decreases completely linearly during the entire observation from 13.8 MSH to 2.7 MSH. Its decay rate is $-1.8$ MSH day$^{-1}$. The total reduced area of each sunspot umbra is 8.1 MSH for S1 and 11.1 MSH for S2, suggesting that the umbral area of sunspot S1 decays more slowly on average than that of sunspot S2. From Figure 3(a2), the penumbral area of sunspot S1 reduces fast from 75.5 MSH to 70.6 MSH at a speed of $-7.2$ MSH day$^{-1}$ at first. Then, it abruptly reduces to 66.8 MSH due to the separation of a part of the penumbra from S1, and it increases to 70.3 MSH as a result of the reappearance of the penumbra in S1. After a tiny area fluctuation, it decreases slowly from 70.3 MSH to 55.7 MSH. The penumbral area of sunspot S2 exhibits a two-stage linear decay pattern. It decreases slowly from 70.2 MSH to

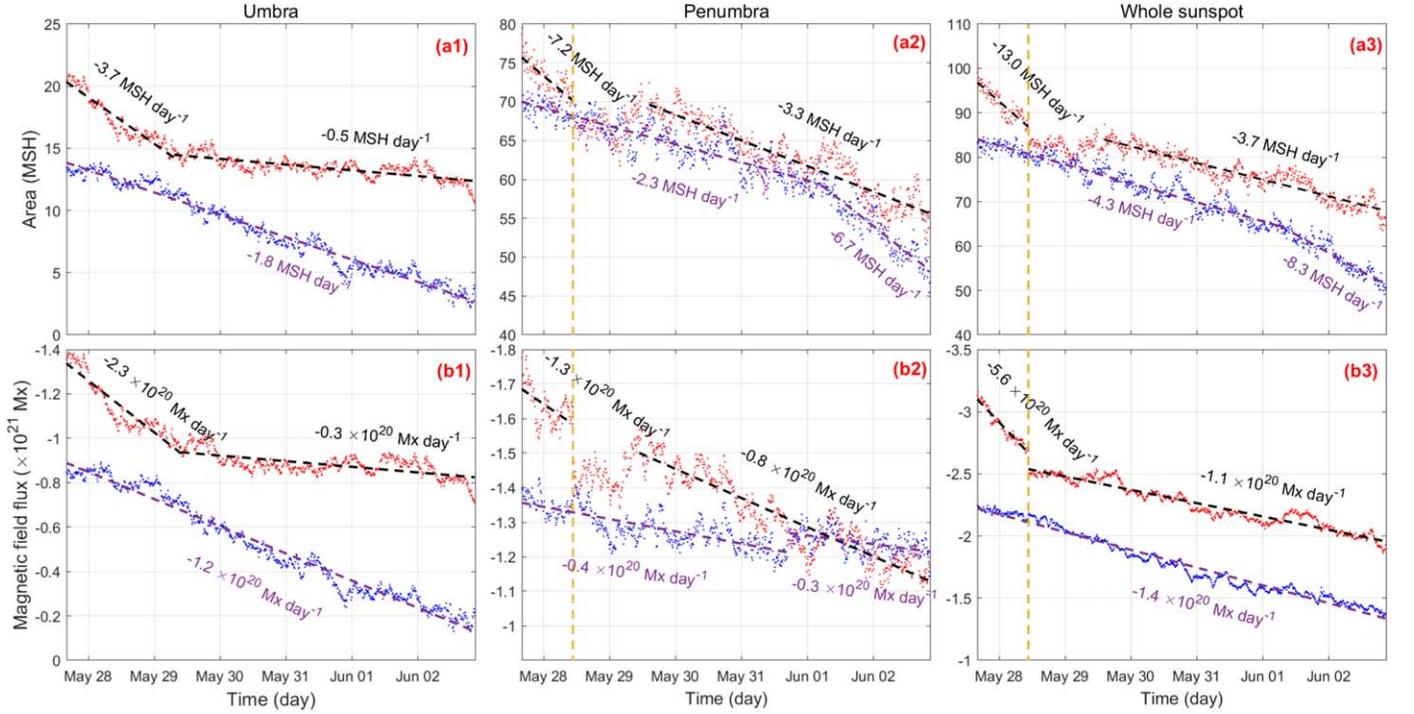

**Figure 3.** Evolution of the area and magnetic flux of the two sunspots. The left, middle, and right panels in the first row display the evolution of the umbral, penumbral, and whole-sunspot area. The red and blue scatter plots represent sunspots S1 and S2 in Figure 1, respectively. The linear model fits the area decay of the sunspot's different components. Its formula is $A(t) = A_0 + DA \cdot (t - t_0)$, where $A(t)$ represents the area (unit: MSH) at time $t$ (unit: days), $t_0$ is the start time of the observation, $A_0$ is the value of each fitting contour at time $t_0$, and $DA$ is the decay rate of the area. The black and purple dashed lines are the fitting contours for the scatter plots of sunspots S1 and S2. The orange dashed lines mark the moment when a large penumbra detaches from sunspot S1. The panels in the second row show the magnetic flux decay of the two sunspots. The abscissa and ordinate represent the time and magnetic flux, and the ordinate is shown in reverse. The color coding of the scatter plots is similar to that of the first row. The linear model is also used to fit the magnetic flux decay of the two sunspots. The formula is $\Phi(t) = \Phi_0 + D\Phi \cdot (t - t_0)$, where $\Phi(t)$ is the magnetic flux at time $t$, $t_0$ is the start time of the observation, $\Phi_0$ is the value of each fitting contour at time $t_0$, and $D\Phi$ is the decay rate of the magnetic flux. The decay rates of the magnetic flux are displayed near the fitting contours.

59.7 MSH, and then reduces rapidly to 48.5 MSH. The decay rates of the two stages are $-2.3$ MSH day$^{-1}$ and $-6.7$ MSH day$^{-1}$, respectively. S1's penumbra decays a little more slowly on average than that of S2. The area decay of each whole sunspot is similar to that of its penumbra. S1 decays fast from 96.5 MSH to 87.3 MSH with a decay rate of $-13$ MSH day$^{-1}$ in the initial stage and then from 84.3 to 68.5 MSH with a decay rate of $-3.7$ MSH day$^{-1}$. The decay of sunspot S2 also exhibits a significant two-stage linear decline; the rates of these stages are $-4.3$ MSH day$^{-1}$ and $-8.3$ MSH day$^{-1}$. The mean decay rates of the two sunspots are $-4.5$ MSH day$^{-1}$ and $-5.2$ MSH day$^{-1}$, and S1 decays more slowly than S2.

Figures 3(b1)–(b3) display the magnetic flux decay of the two sunspots. The magnetic flux variation of each sunspot is fitted by a linear formula: $\Phi(t) = \Phi_0 + D\Phi \cdot (t - t_0)$, where $\Phi(t)$ is the magnetic flux at time $t$, $t_0$ is the start time of the observation, $\Phi_0$ is the value of each fitting contour at time $t_0$, and $D\Phi$ is the decay rate of the magnetic flux. The orange dashed lines mark the moment when a large penumbra detaches from sunspot S1. From Figure 3(b1), the umbral flux of sunspot S1 first declines rapidly from $1.4 \times 10^{21}$ to $0.9 \times 10^{21}$ Mx, then decreases slowly to $0.8 \times 10^{21}$ Mx. The flux decay rates of the two stages are $-2.3 \times 10^{20}$ and $-0.3 \times 10^{20}$ Mx day$^{-1}$, respectively. However, unlike the two stages of S1 umbral flux decay, the umbral flux of sunspot S2 declines linearly at a decay rate of $-1.2 \times 10^{20}$ Mx day$^{-1}$. The penumbral flux of S1 shown in Figure 3(b2) decays at $-1.3 \times 10^{20}$ Mx day$^{-1}$, and decreases suddenly from $1.6 \times 10^{21}$ to $1.4 \times 10^{21}$ Mx at the moment of penumbral detachment. After a tiny fluctuation of the magnetic flux, it reduces slowly from $1.5 \times 10^{21}$ to $1.1 \times 10^{21}$ Mx at a decay rate of $-0.8 \times 10^{20}$ Mx day$^{-1}$. For the penumbral flux of sunspot S2, its two decay rates are $-0.4 \times 10^{20}$ Mx day$^{-1}$ and $-0.3 \times 10^{20}$ Mx day$^{-1}$. Figure 3(b3) shows the magnetic flux decay of the whole sunspot. In contrast to S2, sunspot S1 shows a two-stage linear decay; the decay rates of the two stages are $-5.6 \times 10^{20}$ Mx day$^{-1}$ and $-1.1 \times 10^{20}$ Mx day$^{-1}$, and the mean decay rate is $-1.7 \times 10^{20}$ Mx day$^{-1}$. Sunspot S2 exhibits a linear decay at $-1.4 \times 10^{20}$ Mx day$^{-1}$. The magnetic flux of sunspot S1 on average decays faster than that of sunspot S2.

Figure 4 shows the vector magnetic field evolution for the two sunspots. $B$, $B_t$, $B_z$, and $\gamma$ are the magnetic field strength, the transverse magnetic field strength, the longitudinal magnetic field strength, and the magnetic field inclination angle, respectively. The umbral magnetic field parameters ($B$, $B_t$, $B_z$, and $\gamma$) of sunspot S1 in Figures 4(a1)–(d1) fluctuate around their mean values, which are 2176 G, 799 G, $-2000$ G, and 22°.3, respectively. In contrast to those of sunspot S1, the magnetic field parameters of sunspot S2 all show a slow increase at the beginning, and then decline at steady decay rates. From the blue scatter plot in Figure 4(a1), the parameter $B$ of sunspot S2 increases transiently from 2072 to 2130 G, then reduces steadily to about 1800 G. From Figure 4(d1), the inclination angle of sunspot S2 increases slightly from 20°.5 to 22°, and declines steadily to 15°. In Figures 4(a2) and (b2), it is found that the parameters $B$ and $B_t$ in S1 are higher than those in S2. When compared to those of the umbra, the magnetic parameters of S1's penumbra have similar variations during its decay evolution, and their mean values are 1167 G, 924 G,

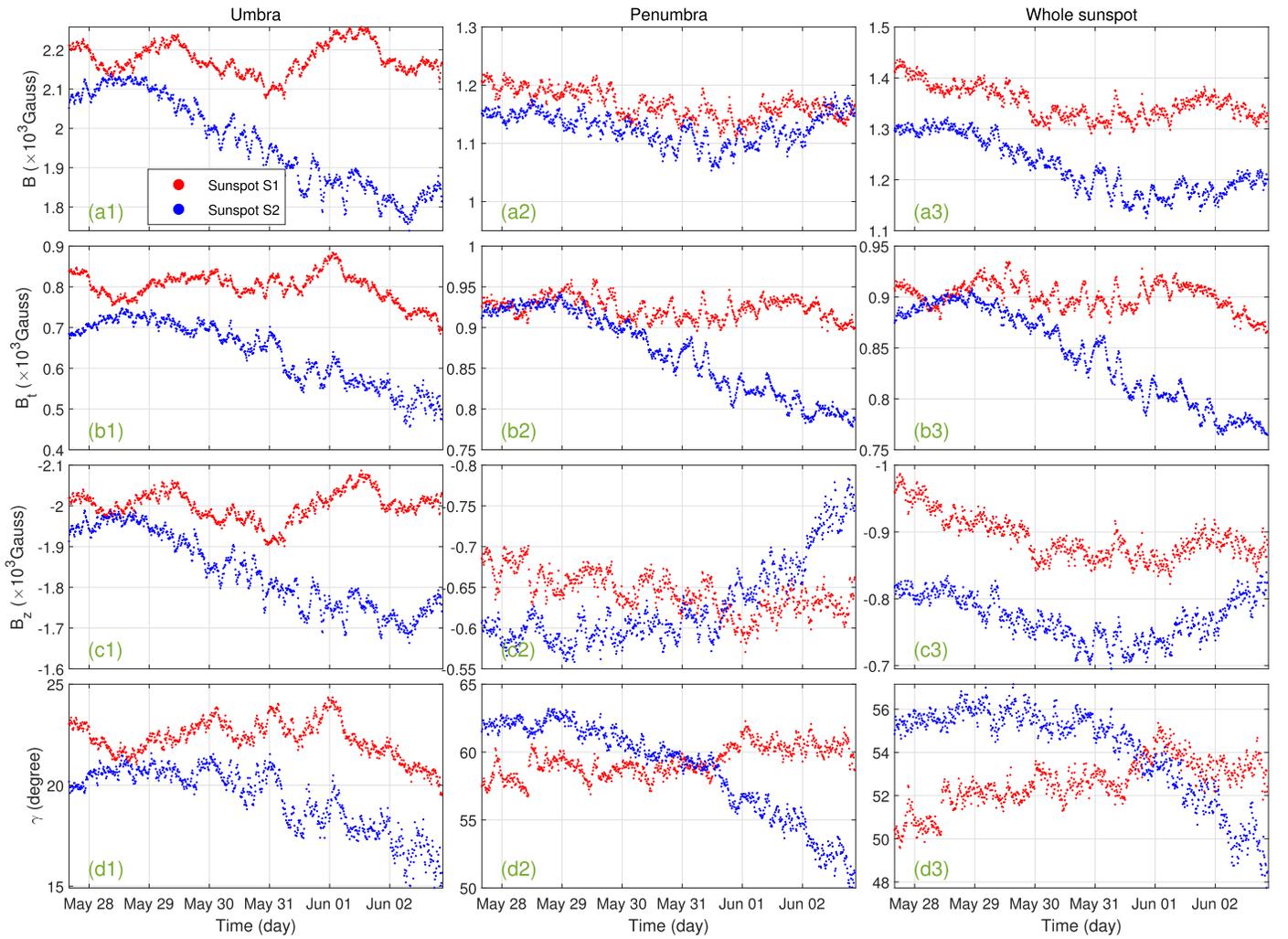

**Figure 4.** Evolution of the vector magnetic field for the two sunspots. The first to fourth rows are for $B$, $B_t$, $B_z$, and $\gamma$, respectively. $B$, $B_t$, $B_z$, and $\gamma$ are the magnetic field strength, transverse magnetic field strength, longitudinal magnetic field strength, and magnetic field inclination angle, respectively. The ordinate of $B_z$ is shown in reverse. Columns (1)–(3) represent the umbra, penumbra, and whole sunspot, respectively. The color coding of the scatter plots is the same as that of Figure 3.

−644 G, and 59°, respectively. On the other hand, the penumbral parameter $B_t$ of sunspot S2 increases slowly from 920 G to 936 G, and reduces steadily to 784 G. The penumbral vertical magnetic field of sunspot S2, after a transient stability, increases fast from 585 G to 760 G. The penumbral inclination angle of sunspot S2 increases slowly at first. Later, it decreases slowly from 63° to 59°, and decreases fast to 51°. The penumbral magnetic field of sunspot S2 tends to be vertical, and the vertical magnetic field increases. The penumbral decay of sunspot S2 is similar to the case studied by Peng et al. (2023). Similar to the area and magnetic flux, the magnetic field parameters of different components within sunspot S2 decline obviously. However, for S1, the umbral and penumbral magnetic field parameters fluctuate approximately around their mean.

### 3.2. The Magnetic Evolution of MMFs

Figure 5 displays the accumulated magnetic flux of the MMFs passing through the ellipse calculated using Equation (4), including that of all the MMFs, just the horizontal ($45° \leqslant \gamma < 90°$) MMFs, and just the vertical ($0° < \gamma < 45°$) MMFs. "Horizontal MMFs" and "vertical MMFs" are hereafter abbreviated as "H-MMFs" and "V-MMFs," respectively. The accumulated magnetic flux of H-MMFs (the magenta curve in Figure 5(a)) increases slowly and then rapidly to $8.8 \times 10^{21}$ Mx with a mean rate of $1.4 \times 10^{21}$ Mx day$^{-1}$. However, in contrast to that of H-MMFs, the accumulated magnetic flux of V-MMFs increases to $1.9 \times 10^{21}$ Mx at $1.1 \times 10^{21}$ Mx day$^{-1}$, slows down, and then increases at a rate of $2.3 \times 10^{20}$ Mx day$^{-1}$. The mean magnetic flux generation rates of the H-MMFs and V-MMFs are $1.4 \times 10^{21}$ Mx day$^{-1}$ and $4.7 \times 10^{20}$ Mx day$^{-1}$, and the mean magnetic flux loss rate of the sunspots is $3.4 \times 10^{20}$ Mx day$^{-1}$. The mean magnetic flux generation rate of the H-MMFs and V-MMFs is 4.1 and 1.4 times higher than the mean magnetic flux loss rate of the sunspots. To compare with the accumulated magnetic flux of V-MMFs, the accumulated magnetic flux loss of the two sunspots is shown by green dots in Figure 5(b). It shows a two-stage linear increase. The loss rate of the rapid stage is $7.1 \times 10^{20}$ Mx day$^{-1}$, then the magnetic flux loss increases slowly at a speed of $2.5 \times 10^{20}$ Mx day$^{-1}$. The first and second red dashed lines in Figure 5(b) mark the moment when a large penumbra detaches from sunspot S1, and the moment when the pore has passed through the reference ellipse, respectively. After 10:48 on May 28, a large penumbra detaches from sunspot S1, and decays rapidly and becomes a pore.

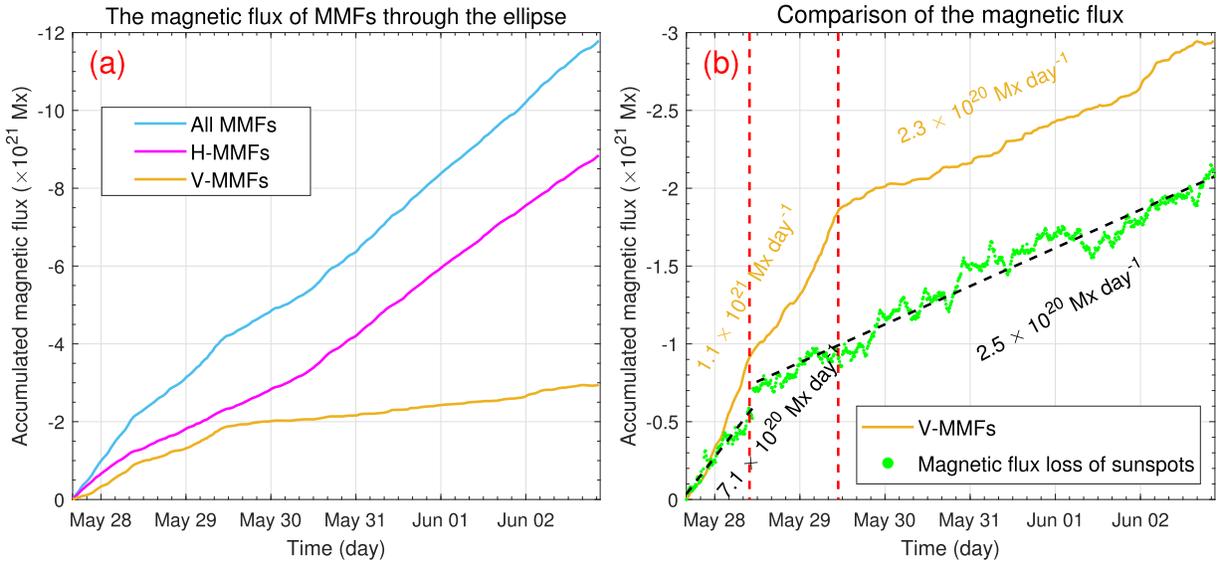

**Figure 5.** Accumulated magnetic flux of MMFs through the ellipse. Panel (a) is the accumulated magnetic flux of MMFs through the ellipse. The blue, magenta, and yellow curves represent the accumulated magnetic flux for all MMFs, H-MMFs ($45° \leqslant \gamma < 90°$), and V-MMFs ($0° < \gamma < 45°$), respectively. Panel (b) is the magnetic flux loss of the two sunspots. $|\Phi(t) - \Phi_0|$ is the magnetic flux loss at time $t$, where $\Phi_0$ is the magnetic flux of the sunspots at the beginning of the observation. The green scatter plot is the magnetic flux loss of the two sunspots, and the fitting method is the same as that in Figure 3. The first and second red dashed lines in panel (b) mark the moment when a large penumbra detaches from sunspot S1, and the moment when the pore has passed through the reference ellipse, respectively.

Meanwhile, the loss rate of the sunspots declines from $7.1 \times 10^{20}$ Mx day$^{-1}$ to $2.5 \times 10^{20}$ Mx day$^{-1}$. After 11:00 on May 29, the pore has passed through the reference ellipse. The magnetic flux generation rate of V-MMFs declines suddenly to $2.3 \times 10^{20}$ Mx day$^{-1}$, and becomes almost equal to the magnetic flux decay rate of the sunspots. This provides a possible evidence to the view that V-MMFs are closely related to the decay of sunspots. The mean magnetic flux generation rate of all MMFs is $1.9 \times 10^{21}$ Mx day$^{-1}$, which is 5.6 times higher than the mean flux-loss rate of the sunspots. These results are in agreement with previous studies (Martínez Pillet 2002; Kubo et al. 2007).

Figure 6 displays the time–distance diagrams of the $B_t$, $B_z$, and $\gamma$ maps along the ellipse slit, whose azimuth angle is described by 0°–360° counterclockwise from the west (the yellow arrow in Figure 1(b)). There is a strong magnetic region from 16:00 on May 27 to about 12:00 on May 29, which is related to the separation of a part of the penumbra from S1. Their inclination angle is less than 45°. Two other strong magnetic field regions are distributed on the ribbons marked by the green dashed curves. They are the V-MMFs from the gap. Accompanying the rotation of the two sunspots, the gap rotates and produces many MMFs whose inclination angle is less than 45°. At other regions, there are no steady signals of V-MMFs. The magnetic flux generated by MMFs from the gap is described in Section 3.3.

### 3.3. The MMFs from the Gap between the Two Sunspots

Figure 7 displays the time–distance diagrams produced by the fan-shaped regions in Figure 1(b). Figures 7(a) and (b) are the time–distance diagrams produced by the upper and lower fan-shaped regions in Figure 1(b), respectively. There are many MMFs from the gap, and several distinct trajectories of the MMFs through the upper and lower fan-shaped regions are marked by the yellow and red dashed lines, respectively. The slopes of the MMFs' trajectories are the velocity of the MMFs. From the figure, the speeds of the MMFs are estimated to be at a range of 0.31–0.58 km s$^{-1}$ with a mean value of 0.4 km s$^{-1}$. The observations conform to previous findings (Brickhouse & Labonte 1988; Zhang et al. 2003; Kubo et al. 2007) that MMFs' motion speed ranges from 0.1 to 1 km s$^{-1}$.

Furthermore, the magnetic flux of the MMFs from the gap is measured though blue arcs (Figure 1(b)), and Figure 8(a) displays the accumulated magnetic flux of all MMFs (the blue curve), the H-MMFs (the magenta curve), and the V-MMFs (the yellow curve). It should be noted that the data before 11:00 UT on May 29 is not included to exclude the disturbance from the detached penumbra. The accumulated flux of V-MMFs increases slowly to $0.5 \times 10^{20}$ Mx at around 18:00 on May 31, and increases fast to $3 \times 10^{20}$ Mx in the end. The mean flux generation rates during the two stages are $0.2 \times 10^{20}$ and $1.2 \times 10^{20}$ Mx day$^{-1}$, respectively. The mean magnetic flux generation rates of H-MMFs for the two stages are $0.8 \times 10^{20}$ and $1.9 \times 10^{20}$ Mx day$^{-1}$, respectively. The mean flux generation rate of H-MMFs is twice that of V-MMFs. The total magnetic flux through the blue arcs is $0.9 \times 10^{21}$ Mx, one-third of which is contributed by V-MMFs. Compared with the magnetic flux from the gap, Figure 8(b) displays the accumulated magnetic flux of the MMFs passing through the ellipse except for two arcs (called the nongap region). The total accumulated magnetic flux of V-MMFs and H-MMFs is $0.8 \times 10^{21}$ Mx and $6 \times 10^{21}$ Mx, respectively. The accumulated flux of H-MMFs is 7.5 times higher than that of V-MMFs, implying that almost all of the magnetic flux of the MMFs flowing out of the nongap region is contributed by H-MMFs.

### 4. Conclusions and Discussion

To better understand the relationship between the decay of sunspots and MMFs, the evolution of NOAA AR 13023 is studied. A gap exists between the two adjacent sunspots where the penumbra is absent. The magnetic flux and its variations generated by MMFs around the sunspots and MMFs from the gap are measured. These MMFs are classified as H-MMFs and V-MMFs. The main results are listed as follows:

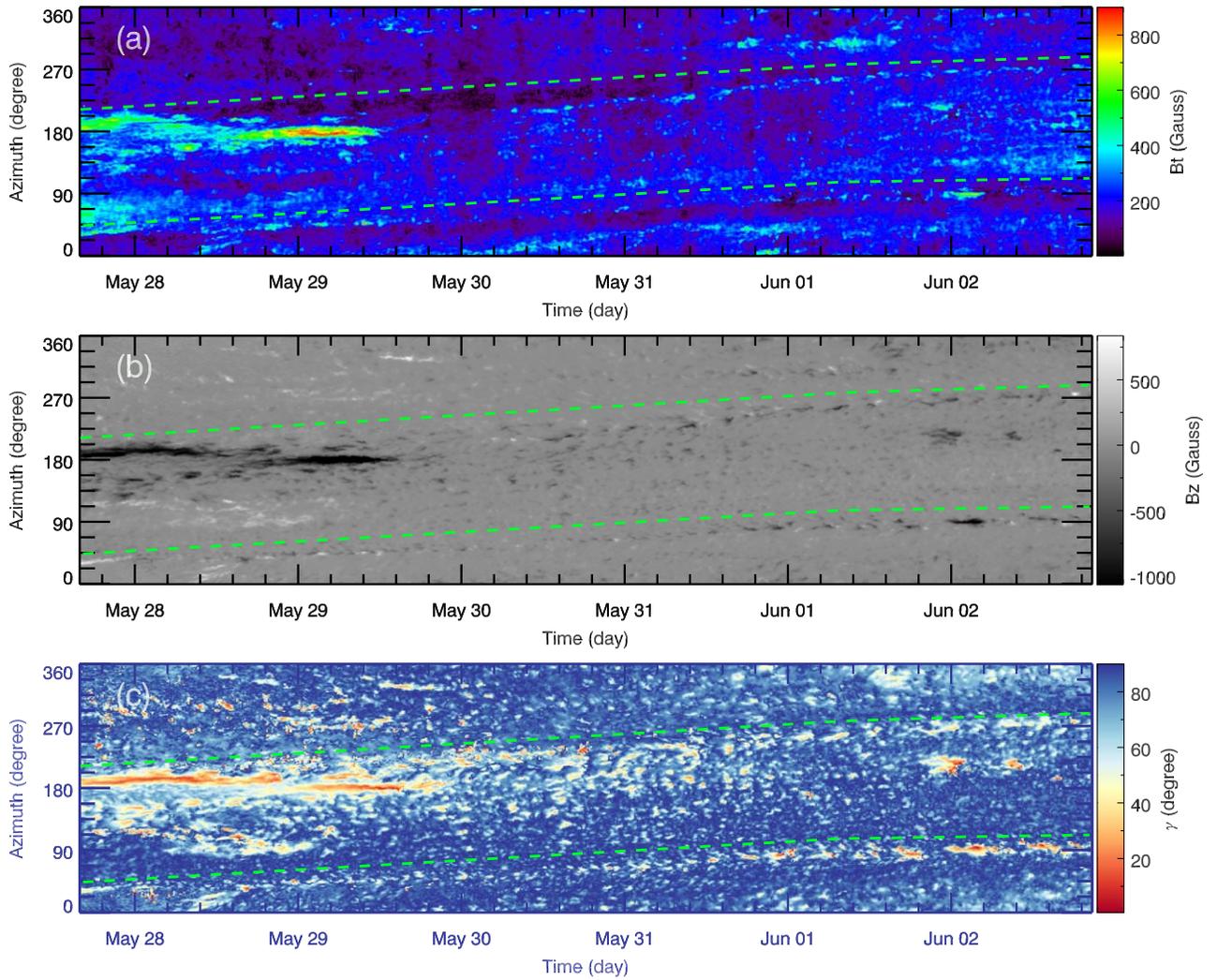

**Figure 6.** Vector magnetic field distribution of the MMFs passing through the ellipse with the azimuth angle. The upper, middle, and lower panels are the time–distance diagrams of the $B_t$, $B_z$, and $\gamma$ maps along the ellipse slit. The meanings of $B_t$, $B_z$, and $\gamma$ are the same as those of Figure 2. The azimuth angle is determined by 0°–360° counterclockwise from the west (the yellow arrow in Figure 1(b)). The green dashed lines mark the tracks of the MMFs from the gap.

(1) The mean decay rates of the area of sunspots S1 and S2 are −4.5 and −5.2 MSH day$^{-1}$, respectively. The mean decay rates of their magnetic flux are $-1.7 \times 10^{20}$ and $-1.4 \times 10^{20}$ Mx day$^{-1}$, respectively. The magnetic flux of sunspot S1 decays faster than that of sunspot S2.

(2) Similar to the area and magnetic flux, the magnetic field parameters of different components within sunspot S2 decline obviously. The penumbral magnetic field of sunspot S2 becomes more vertical during its decay. However, for S1, the umbral and penumbral magnetic field parameters fluctuate approximately around their mean.

(3) The generation rate of the magnetic flux of all MMFs including the separated pore is on average 5.6 times the flux-loss rate of the sunspots. The magnetic flux generation rate of V-MMFs including the separated pore is on average 1.4 times the magnetic flux loss rate of the sunspots. After the pore passes through the ellipse, the magnetic flux generation rate of V-MMFs is almost similar to the magnetic flux loss rate of the two sunspots.

(4) The speeds of the MMFs from the gap range from 0.3 to 0.6 km s$^{-1}$ with a mean speed of 0.4 km s$^{-1}$.

(5) The total accumulated magnetic flux of V-MMFs from the gap is one-third of the total magnetic flux of all MMFs from the gap. However, at the position of the nongap region, few V-MMFs stream out from the sunspots and their magnetic flux is close to one-ninth of that of all MMFs from the nongap region.

The areas of the two sunspots show a two-stage linear decay. The mean area decay rates for sunspots S1 and S2 are −4.5 and −5.2 MSH day$^{-1}$, respectively. These results are lower than the decay rates of $\alpha$-configuration sunspots obtained by Li et al. (2021). Meanwhile, the magnetic flux decay of sunspot S1 also exhibits a two-stage linear pattern, and its mean rate is $-1.7 \times 10^{20}$ Mx day$^{-1}$. However, the magnetic flux of sunspot S2 always decays linearly with a decay rate of $-1.4 \times 10^{20}$ Mx day$^{-1}$. In terms of the magnetic flux, sunspot S1 decays faster than sunspot S2. The magnetic flux decay rates of the two sunspots are slightly less than the typical decay rate of $2-4 \times 10^{20}$ Mx day$^{-1}$ (Kubo et al. 2008a; Sheeley et al. 2017). The turbulent diffusion of the sunspot magnetic field shows the linear decay law of the area (Meyer et al. 1974; Krause & Rudiger 1975). The magnetic field crosses the whole area of the sunspot and diffuses at a constant diffusivity. In our research, the total area and magnetic flux of sunspot S1 almost show a two-stage linear decay. The magnetic diffusion is divided into two stages, namely, fast and slow stages. That is similar to the eight α-configuration sunspots' decay patterns studied by Li et al. (2021). Similar to the area and magnetic flux, the magnetic field parameters of different components within sunspot S2 decline obviously. However, for S1, the umbral and penumbral magnetic field parameters fluctuate approximately around their mean. For sunspot S2, the magnetic field of the penumbra becomes vertical, and the mean $B_z$ of the penumbra increases. This result is similar to those of previous studies (Wang et al. 2004; Watanabe et al. 2014). There are some possible interpretations, for instance, the submergence of the horizontal magnetic field in the penumbra (Peng et al. 2023), the reconstruction of the magnetic field caused by flares (Wang et al. 2012), or flux emergence (Verma et al. 2018). There are no flares near the sunspots, and no flux emergence around sunspot S2. With the submergence of the horizontal magnetic field in the penumbra, the relatively vertical magnetic field is left, resulting in the reduction of the mean inclination angle in the penumbra. For sunspot S1, it is not observed that the magnetic field of the penumbra becomes more vertical.

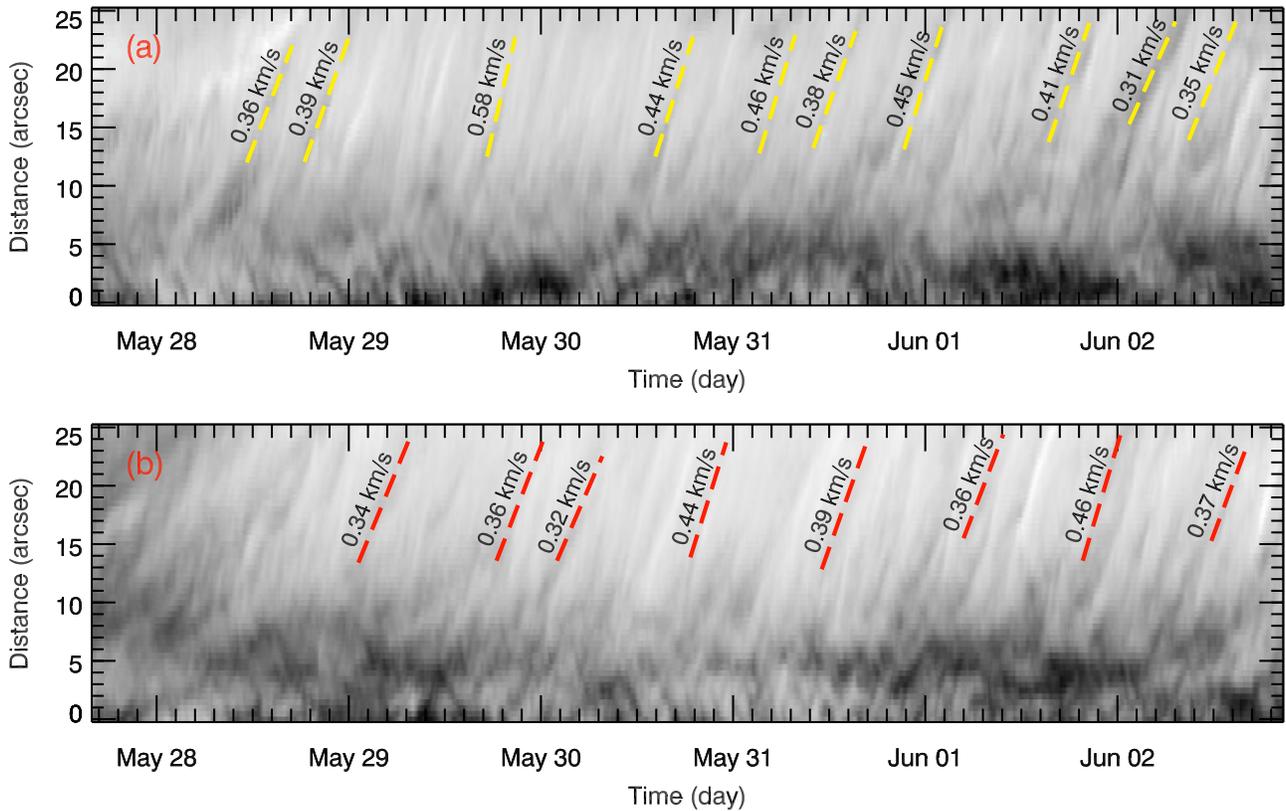

**Figure 7.** The time–distance diagrams produced by the fan-shaped regions in Figure 1(b). Panels (a) and (b) are the time–distance diagrams produced by the upper and lower fan-shaped regions in Figure 1(b), respectively. The horizontal and vertical axes of each panel represent time (unit: days) and distance (unit: arcseconds). The yellow and red dashed lines mark several distinct trajectories of MMFs through the upper and lower fan-shaped regions, respectively. The slopes of the MMFs' trajectories are the velocity of the MMFs, and the corresponding velocities are listed near the dashed lines.

Martínez Pillet (2002) found that MMFs' flux generation rates are 3–8 times greater than the magnetic flux loss rates of sunspots, and concluded that only Type II MMFs are related to the diffusion of the sunspot magnetic flux. Thomas et al. (2002) and Weiss et al. (2004) suggested that Type I and Type III MMFs correspond to the intersections of the horizontal magnetic field extended from the penumbra with the solar surface, and Type II MMFs are detached from the vertical components of the penumbral interlocking-comb structure and thus are responsible for the decay of sunspots. Kubo et al. (2007) found that MMFs detached from the penumbral spine (vertical ones) have vertical magnetic fields with the same polarity as the parent sunspots, and the magnetic flux carried by V-MMFs is 1–3 times higher than the magnetic flux loss of the sunspots. They concluded that isolated V-MMFs are only related to the disintegration of the sunspot. In our study, the mean magnetic flux generation rate of all MMFs including the separated pore is 5.6 times higher than the mean magnetic flux loss rate of the sunspots, and the H-MMFs' and V-MMFs' flux generation rates are on average 4.1 and 1.4 times higher than the magnetic flux loss rate of the sunspots. These results are consistent with the previous results of Kubo et al. (2007) and Martínez Pillet (2002), and show that only a small fraction of MMFs are related to the decay of sunspots. Furthermore, after about 11:00 UT on May 29, when the pore passes through the ellipse completely and no longer affects the calculation of the magnetic flux of MMFs, the magnetic flux loss rate of the sunspots is almost similar to the magnetic flux generation rate of V-MMFs. This provides a possible evidence that the MMFs with vertical magnetic fields are responsible for the decay of the sunspots.

The magnetic flux of the MMFs (vertical or horizontal) from the gap and nongap regions shows that the accumulated magnetic flux of the V-MMFs from the gap is one-third of the total magnetic flux. However, for the V-MMFs from the nongap region, their accumulated magnetic flux is only about one-ninth of the total magnetic flux. The sunspots at the gap lack their penumbra, and the MMFs at the gap may come from the umbra, and most of them have more vertical magnetic field. In contrast, at the nongap region, a majority of the MMFs are relevant to the penumbra. As a result, MMFs can also be produced directly from the sunspot umbra, rather than strictly from the penumbra (Thomas et al. 2002; Sainz Dalda & Martínez Pillet 2005; Cabrera Solana et al. 2006; Kubo et al. 2008b). This conclusion still needs more observation cases to be verified; in particular, the MMFs from pores can be studied. The statistical study of the decay of sunspots whose partial umbra is naked can further confirm this view. Further study of MMFs around pores can also provide possible demonstrations.

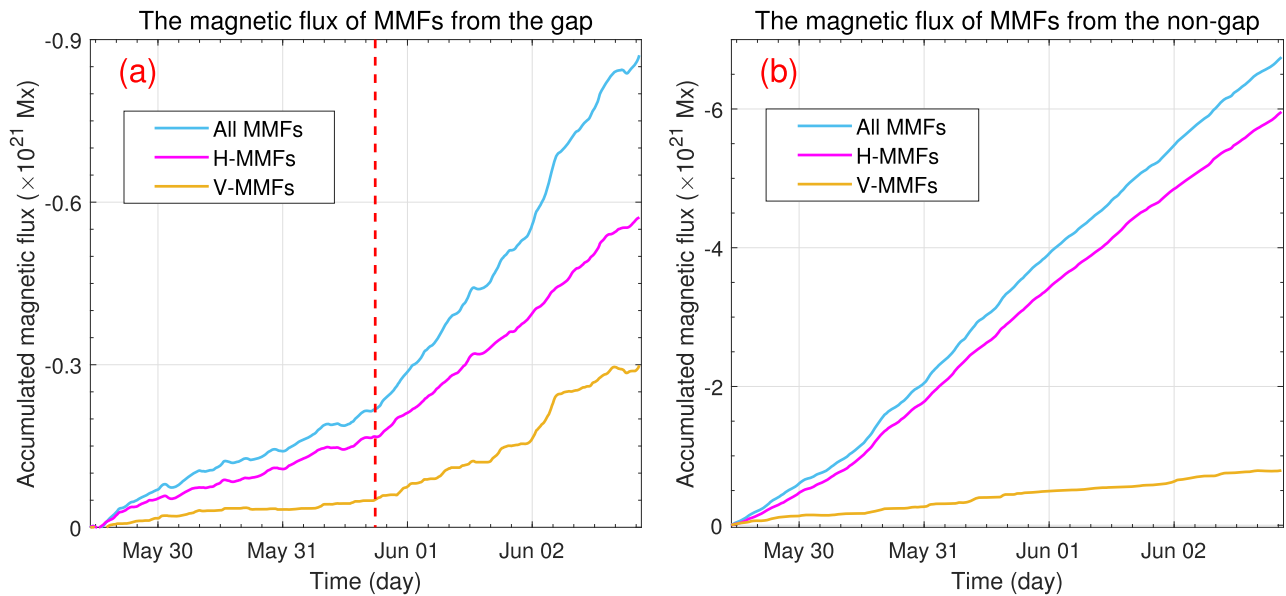

**Figure 8.** Accumulated magnetic flux of the MMFs flowing out of the gap. The blue, magenta, and yellow curves represent the accumulated magnetic flux for all MMFs, H-MMFs ($45° \leqslant \gamma < 90°$), and V-MMFs ($0° < \gamma < 45°$), respectively. For comparison with panel (a), the accumulated flux of the MMFs from the nongap region is calculated in panel (b). The red dashed line in panel (a) marks the moment when the increases of the three curves become fast.

### Acknowledgments

We wish to express our gratitude to the anonymous referee for constructive comments and suggestions, and thank the SDO/HMI teams for the AR data support. We are grateful to Xiaoli Yan and Zhi Xu for their fruitful comments and suggestions. This work is supported by the Strategic Priority Research Program of the Chinese Academy of Sciences, grant No. XDB0560000, the National Natural Science Foundation of China (11973084, 11973088, 12003066, 11527804, U1831210, 12203097, 12325303, and 12003064), the Yunnan Key Laboratory of Solar Physics and Space Science under No. 202205AG070009, the Yunnan Science Foundation of China under Nos. 202101AT070032, 202201AT070194, and 202301AT070347, the Youth Innovation Promotion Association, the Chinese Academy of Sciences (CAS) (No. 2019061), the CAS Light of West China Program, the Yunnan Science Foundation for Distinguished Young Scholars No. 202001AV070004, the Key Research and Development Project of Yunnan Province under No. 202003AD150019, the Young Elite Scientists Sponsorship Program by YNAST.9, the Yunnan Fundamental Research Projects (grant No. 202301AT070349), and a grant associated with the Project of the Group for Innovation of Yunnan Province.

### ORCID iDs

Yang Peng ⓘ https://orcid.org/0000-0002-0492-6789
Zhike Xue ⓘ https://orcid.org/0000-0002-6526-5363
Zhongquan Qu ⓘ https://orcid.org/0000-0002-1787-325X
Jincheng Wang ⓘ https://orcid.org/0000-0003-4393-9731
Zhe Xu ⓘ https://orcid.org/0000-0002-9121-9686
Liheng Yang ⓘ https://orcid.org/0000-0003-0236-2243